\documentclass[a4paper,11pt]{article}
\title{\bf An Integer Programming approach to the\\Hospital/Residents problem with Ties}
\author{Augustine Kwanashie and David F. Manlove \thanks{Supported by Engineering and Physical Sciences Research Council grant GR/EP/K010042/1.}\\ \\{\small \emph{School of Computing Science, University of Glasgow, UK}} \\{\small \tt{a.kwanashie.1@research.gla.ac.uk}, \tt{David.Manlove@glasgow.ac.uk}}}

\date{}

\usepackage{makeidx}
\usepackage{hyperref}
\usepackage{pdflscape}
\usepackage{amsmath}
\usepackage{amsfonts}
\usepackage{verbatim}
\usepackage{graphicx}
\usepackage{algorithm}
\usepackage{algorithmic}
\usepackage{textcomp}
\usepackage{mathtools}
\usepackage{float}
\usepackage{epstopdf}

\newtheorem{lemma1}{Lemma}[section]
\newtheorem{theorem1}[lemma1]{Theorem}

\newenvironment{proof}[1][Proof]{\begin{trivlist}
\item[\hskip \labelsep {\bfseries #1}]}{\end{trivlist}}

\begin{document}
\setcounter{page}{1}
\bibliographystyle{plain}
\maketitle
\pagenumbering{arabic}

\begin{abstract}
The classical Hospitals/Residents problem (HR) models the assignment of junior doctors to hospitals based on their preferences over one another. In an instance of this problem, a stable matching $M$ is sought which ensures that no blocking pair can exist in which a resident $r$ and hospital $h$ can improve relative to $M$ by becoming assigned to each other. Such a situation is undesirable as it could naturally lead to $r$ and $h$ forming a private arrangement outside of the matching.  
 
The original HR model assumes that preference lists are strictly ordered. However in practice, this may be an unreasonable assumption: an agent may find two or more agents equally acceptable, giving rise to \emph{ties} in its preference list. We thus obtain the Hospitals/Residents problem with Ties (HRT). In such an instance, stable matchings may have different sizes and MAX HRT, the problem of finding a maximum cardinality stable matching, is NP-hard.
 
In this paper we describe an Integer Programming (IP) model for MAX HRT. We also provide some details on the implementation of the model. Finally we present results obtained from an empirical evaluation of the IP model based on real-world and randomly generated problem instances. 
\end{abstract}

\section{Introduction}
The Hospital Residents Problem (HR) has applications in a number of centralised matching schemes which seek to match graduating medical students (residents) to hospital positions. Examples of such schemes include the Scottish Foundation Allocation Scheme (SFAS) \cite{SFAS} and the National Resident Matching Program (NRMP) \cite{NRMP} in Scotland and the US respectively. The challenges presented by these and other applications have motivated research in the area of algorithms for matching problems. 

Formally an instance $I$ of HR involves a set  $R=\{r_1, r_2, ... ,$ $r_{n_1}\}$ of \emph{residents} and  $H=\{h_1, h_2, ... , h_{n_2}\}$ of \emph{hospitals}. Each resident $r_i \in R$ ranks a subset of $H$ in strict order of preference with each hospital $h_j \in H$ ranking a subset of $R$, consisting of those residents who ranked $h_j$, in strict order of preference. Each hospital $h_j$ also has a capacity $c_j \in \mathbb{Z}^+$  indicating the maximum number of residents that can be assigned to it. A pair $(r_i, h_j)$ is called an \emph{acceptable pair} if $h_j$ appears in $r_i$'s preference list and $r_i$ on $h_j$'s preference list.  A \emph{matching} $M$ is a set of acceptable pairs such that each resident is assigned to at most one hospital and the number of residents assigned to each hospital does not exceed its capacity.  A resident $r_i$ is \emph{unmatched} in $M$ if no acceptable pair in $M$ contains $r_i$. We denote the hospital assigned to resident $r_i$ in $M$ as $M(r_i)$ (if $r_i$ is unmatched in $M$ then $M(r_i)$ is undefined) and the set of residents assigned to hospital $h_j$ in $M$ as $M(h_j)$.  A hospital $h_j$ is \emph{under-subscribed} in $M$ if  $|M(h_j)| < c_j$. An acceptable pair $(r_i, h_j)$ can \emph{block} a matching $M$ or forms a \emph{blocking pair} with respect to $M$ if $r_i$ is either unmatched or prefers $h_j$ to $M(r_i)$ and $h_j$ is either under-subscribed or prefers $r_i$ to at least one resident in $M(h_j)$. A matching $M$ is said to be \emph{stable} if there exists no blocking pair with respect to $M$. 

We consider a  generalisation of HR which occurs when the preference lists of the residents and hospitals are allowed to contain \emph{ties}, thus forming the Hospital/Residents Problem with Ties (HRT). In an HRT instance a resident (hospital respectively) is indifferent between all hospitals (residents respectively) in the same tie on its preference list. In this context various definitions of stability exists. We consider \emph{weak stability} \cite{irv94} in which a pair $(r_i, h_j)$ can \emph{block} a matching $M$ if $r_i$ is either unmatched or strictly prefers $h_j$ to $M(r_i)$ and $h_j$ is either under-subscribed or strictly prefers $r_i$ to at least one resident in $M(h_j)$. A matching $M$ is said to be \emph{weakly stable} if there exists no blocking pairs with respect to $M$.  Henceforth we will refer to a weakly stable matching as simply a stable matching. Figure \ref{hr_sample_instance} shows a sample HRT instance.

Every instance of the HRT problem admits at least one weakly stable matching. This can be obtained by breaking the ties in both sets of preference lists in an arbitrary manner, thus giving rise to a HR instance which can then be solved using the Gale-Shapley algorithm for HR \cite{GS62}. The resulting stable matching is then stable in the original HR instance. However, in general, the order in which the ties are broken yields stable matchings of varying sizes \cite{MIIMM02} and the problem of finding a maximum weakly stable matching given an HRT instance (MAX HRT) is known to be NP-hard \cite{MIIMM02}. For example, the instance in Figure \ref{hr_sample_instance}  admits the following stable matchings: $M_0 = \{(r_1, h_1), (r_2, h_1), (r_3, h_3),$ $(r_5, h_2), (r_6, h_2)\}$ and $M_1 = \{(r_1, h_1), (r_2, h_1), (r_3, h_3),  (r_4, h_2), (r_5, h_3), (r_6, h_2)\}$ of  sizes 5 and 6 respectively. Various approximation algorithms for MAX HRT can be found in the literature \cite{McD09, ZK12} with the best current algorithm achieving a performance guarantee of $3/2$.

Due to the NP-hardness of MAX HRT and the need to maximize the cardinality of stable matchings in practical applications, \emph{Integer Programming} (IP) can be used to solve MAX HRT instances to optimality. This paper presents a new IP model for MAX HRT (Section \ref{compact}).  In Section \ref{imp} we provide some details on the implementation of the model. Section \ref{eval} summarises some of the results obtained by evaluating the model against real-world and randomly generated problem instances. Finally Section \ref{future} highlights some interesting open problems. 

\begin{figure}[htb]
\centering
 \begin{minipage}{150pt}
\begin{align*}
&\mbox{residents' preferences} &\>\>\>\>\>\>\>\>\>\>\>\>&\mbox{hospitals' capacities and preferences}\\
&r_1: h_1\>\>\>h_2  &&h_1: (2): r_1\>\>\>r_2\>\>\>r_3\>\>\>r_6 \\
&r_2: h_1  &&h_2: (2): r_2\>\>\>r_1\>\>\>r_6\>\>\>(r_4\>\>\>r_5) \\
&r_3: h_1\>\>\>h_3  &&h_3: (2): r_5\>\>\>r_3 \\
&r_4: h_2 \\
&r_5: h_2\>\>\>h_3 \\
&r_6: h_1\>\>\>h_2
\end{align*}
 \end{minipage}
\caption{HRT instance $I$ (entries in round brackets are tied)}
\label{hr_sample_instance}
\end{figure}

\section{Previous linear, integer  and constraint programming models}
\label{survey}
Various Linear Programming (LP) formulations for stable matching problems appear in the literature. The notable LP formulations for the one-to-one variant of HR, the Stable Marriage problem (SM) (the restriction of HR in which $n_1=n_2$ and $c_j=1$ for all $h_j \in H$ and every resident finds every hospital acceptable), are presented by Gusfield and Irving \cite{GI89} and Vande Vate \cite{VV89}. Extensions of the Vande Vate SM model to the Stable Marriage problem with Incomplete lists (SMI) (the restriction of HR in which $c_j=1$ for all $h_j \in H$) and HR problems were given by Rothblum \cite{Rot92}. LP formulations for SM have also been extended to the case where we seek an egalitarian, minimum regret or minimum weight stable matchings \cite{GI89, VV89, TS98}. Matching problems have also been modelled as constraint satisfaction problems. A constraint programming approach to HR  was presented in \cite{MOPU05} and an extension to HRT presented in \cite{Uns08}. A review of CSP and LP approaches to stable matching problems can be found in \cite{DM12}.

IP formulations add the restriction that the variables involved  must all have integral values in any feasible solution. Although in general, the integer programming problem is NP-complete \cite{Kar72}, a wide variety of practical problems can be formulated and solved using integer programming.  An IP formulation for MAX SMTI was presented by Podhradsk\`{y} \cite{POD11} where he compared the sizes of weakly stable matchings obtained by running various approximation algorithms on SMTI instances against the maximum weakly stable matchings obtained from his IP formulation. In this paper we extend Podhradsk\`{y}'s IP model to the HRT case.

\section{An IP model for MAX HRT}
\label{compact}
In this section we describe an  IP model for MAX HRT. Let $I$ be an instance of HRT consisting of a set $R = \{r_1, r_2, ... , r_{n_1}\}$ of residents and $H = \{h_1, h_2, ... , h_{n_2}\}$ of hospitals. We denote the binary variable $x_{i,j} ~(1 \leq i \leq n_1, 1 \leq j \leq n_2)$ to represent an acceptable pair in $I$ formed by resident $r_i$ and hospital $h_j$. Variable $x_{i,j}$ will indicate whether $r_i$ is matched to $h_j$ in a solution or not: if $x_{i,j}=1$ in a given solution $J$ then $r_i$ is matched to $h_j$ in $M$ (the matching obtained from $J$), else $r_i$ is not matched to $h_j$ in $M$. We  define $rank(r_i, h_j)$, the rank of $h_j$ on $r_i$'s preference list to be $k + 1$ where  $k$ is the number of hospitals that $r_i$ strictly prefers to $h_j$. An analogous definition for $rank(h_j, r_i)$ holds. Obviously for HRT instances agents in the same tie have the same rank. We define $rank(r_i, h_j) = rank(h_j, r_i) = \infty$ for an unacceptable pair $(r_i, h_j)$. With respect to a pair $(r_i, h_j)$, we define the set  $T_{i,j}=\{r_{p} \in R: rank(h_j,r_{p}) \leq rank(h_j, r_i)\}$ and $S_{i,j}=\{h_{q} \in H: rank(r_i,h_{q}) \leq rank(r_i, h_j)\}$. We also define the set $P(r_i)$ to be the set of hospitals that $r_i$ finds acceptable and $P(h_j)$ to be the set of residents that $h_j$ finds acceptable. Figure \ref{model4} shows the resulting model. Constraint 1 ensures that each resident is matched to at most one hospital and Constraint 2 ensures that each hospital does not exceed its capacity. Finally Constraint 3 ensures that the matching is stable by ruling out the existence of any blocking pair.

\begin{figure}[htb]
\centering
\fbox{
\begin{minipage}{400pt}
\begin{align*}
&\max \sum\limits_{i=1}^{n_1} \sum\limits_{h_j \in P(r_i)} x_{i,j}\\ \\
&\mbox{subject  to} \\
&1. \>\>\>\sum\limits_{h_j \in P(r_i)} x_{i,j}  \leq 1 & &(1 \leq i \leq n_1) \\
&2. \>\>\>\sum\limits_{r_i \in P(h_j)} x_{i,j} \leq  c_j & &(1 \leq j \leq n_2) \\
&3. \>\>\>c_j \left( 1 - \sum\limits_{h_{q} \in S_{i,j}} x_{i,q} \right) -  \sum\limits_{r_{p} \in T_{i,j}} x_{p,j}\leq 0 & & (1 \leq i \leq n_1, h_j \in P(r_i)) \\
& x_{i,j} \in \{0,1\}
\end{align*}
 \end{minipage}
}
\caption{{\bf model1}: A HRT IP model}
\label{model4}
\end{figure}

\begin{theorem1}
\label{model1proof}
Given a HRT instance $I$ modeled as an IP using model1, a feasible solution to  model1 produces a weakly stable matching in $I$. Conversely a weakly stable matching in $I$ corresponds to a feasible solution to  model1.
\end{theorem1}

\begin{proof}
Assume that the IP model has a feasible solution involving variables $x_{i,j} $ for each $ i~(1 \leq i \leq n_1),~ h_j \in P(r_i)$. Let  $M = \{(r_i, h_j) \in R\times H: r_i \in R \wedge h_j \in P(r_i) \wedge x_{i,j}=1\}$ 
be the set of pairs in $I$ generated from the IP solution. By constraints 1 and 2, $M$ is a matching: each resident is matched to at most one hospital, residents are only ever assigned to acceptable hospitals, and hospitals do not exceed their capacities. Constraint 3 ensures that the matching is weakly stable: if $M$ was not a weakly stable matching then some pair $(r_i, h_j)$ would block $M$. Thus $r_i$ prefers $h_j$ to $M(r_i)$ or $r_i$ is unmatched. In both cases, $\sum_{h_{q} \in S_{i,j}} x_{i,q}=0$, meaning the first term in constraint 3 would yield a value of $c_j$.  Also $h_j$ is either under-subscribed or $h_j$ strictly prefers $r_i$ to one of the residents in $M(h_j)$. In both cases, the second term of constraint 3 would be less than $c_j$ thus the IP solution would be infeasible, a contradiction.

Conversely, assume that $M$ is a weakly stable matching in $I$. We can show that $M$ produces a feasible solution to model1. Initially for all $i~(1 \leq i \leq n_1),~j~($such that $h_j \in P(r_i))$, let $x_{i,j}=0$. If $(r_i, h_j) \in M$ then set $x_{i,j}=1$. Constraints 1 and 2 are satisfied as $M$ is a matching. For constraint 3  not to be satisfied the first term of the left-hand-side must be greater than the second. Thus for some $i~(1 \leq i \leq n_1)$ and $j~(h_j \in P(r_i))$, $r_i$ is either unmatched or strictly prefers $h_j$ to $M(r_i)$, thus making the first term result in a value of $c_j$ as the first term can only have a value of 0 or $c_j$. The second term of the left-hand-side must then be less than $c_j$. This means that the number of residents $r_p$ assigned to $h_j$, such that $h_j$ either prefers $r_p$ to $r_i$ or is indifferent between them, is less than $c_j$. Thus $(r_i, h_j)$ would block $M$, a contradiction.
\end{proof}

\section{Implementing the model}
\label{imp}
In this section we describe some techniques used to reduce the size of the HRT model generated and improve the performance of the IP solver. We also specify how the implementation was tested for correctness.

\subsection{Reducing the model size}
 Techniques were described in \cite{IM09} for removing acceptable pairs that cannot be part of any stable matching from HRT instances with ties on one side of the preference lists only. The \emph{hospitals-offer} and \emph{residents-apply} algorithms described therein identify pairs that cannot be involved in any stable matching, nor form a blocking pair with respect to any stable matching, and remove them from the instance.  This produces a reduced HRT instance that would yield fewer variables and constraints when modelled as an IP, thus speeding up the optimisation process. The original instance and the reduced instance have the same set of stable matchings. These techniques are only applicable to instances where ties exists on one side only. This is a natural restriction that can be found in practice where hospitals usually rank residents based on performance grades (which can easily be tied) with residents being required to provide a strict preference of a bounded length (say 5 hospitals). 

To help explain the hospitals-offer and residents-offer procedures, we will define some useful terms. We begin with the   hospitals-offer procedure shown in Algorithm \ref{hosp-off}. The process involves hospitals offering positions to residents, who form provisional arrangements with the hospitals.  We define an \emph{active tie} $T_j$ on a hospital $h_j$'s preference list, as a tie containing one or more residents that immediately follows the least preferred resident currently assigned to $h_j$. Quantity $t_j = |T_j|$ may be $0$ if such a tie does not exist. Also the value of $t_j$ may change during the execution of the hospitals-offer algorithm due to potential deletions of pairs from the problem instance. We denote the number of vacancies $v_j$ in $h_j$ as the difference between the capacity $c_j$ and the number of residents currently assigned to it. Hospital $h_j$ is full when $v_j = 0$ and undersubscribed when $v_j > 0$. For each hospital that hasa number of  vacancies greater or equal to the size of the current active tie, the residents in the active tie are assigned to that hospital. If these residents were previously matched to other hospitals, those assignments are broken. For each of these residents $r_i \in T_j$, this is followed by the removal, from $r_i$'s preference list, of all the hospitals that are successors to $h_j$.  The process terminates when each hospital has an active tie of length 0, or its number of vacancies is less than the size of its active tie.

\begin{algorithm}[htb]
\small
\caption{Hospitals-offer}
\label{hosp-off}
\begin{algorithmic}[1]
\WHILE{(there is a hospital $h_j$ such that $v_j \geq t_j > 0$)}
	\FOR{(each resident $r_i$ in $t_j$)}
		\IF{($r_i$ is already assigned, say to $h_k$)}
			\STATE unassign $r_i$;
			\STATE increment $v_k$;
		\ENDIF
		\STATE assign $r_i$ to $h_j$;
		\STATE decrement $v_j$;
		\FOR{(each hospital $h_k$ that is a strict successor of $h_j$ on $r_i$'s list)}
			\STATE delete the pair ($r_i, h_k$) from the preference lists;
		\ENDFOR
	\ENDFOR
\ENDWHILE
\end{algorithmic}
\normalsize
\end{algorithm}

\begin{algorithm}[htb]
\small
\caption{Residents-apply}
\label{res-off}
\begin{algorithmic}[1]
\WHILE{(some resident $r_i$ is free and has a nonempty list)}
	\STATE $h_j$ = first hospital on $r_i$'s list;
	\STATE assign $r_i$ to $h_j$;
	\STATE increment $a_j$;
	\IF{($a_j \geq c_j$)}	
		\STATE let $r_k$ be one of $h_j$'s $c^{th}$-choice assignees where $c= c_j$;
		\FOR{(each strict successor $r_l$ of $r_k$ in $h_j$'s list)}
			\IF{($r_l$ is assigned to $h_j$)}
				\STATE break the assignment;
				\STATE decrement $a_j$;
			\ENDIF
			\STATE delete pair $(r_l, h_j)$;
		\ENDFOR
	\ENDIF
\ENDWHILE
\end{algorithmic}
\normalsize
\end{algorithm}

The residents-apply procedure shown in Algorithm \ref{res-off}, involves residents successfully applying to hospitals on their preference list in preference order, again forming provisional assignment with the hospitals. We define $a_j$ to be the the number of assignees of a hospital $h_j$. Each unassigned resident (say $r_i$) applies to the first hospital on his/her preference list (say $h_j$). Resident $r_i$ is then temporarily assigned to $h_j$ and $a_j$ is incremented. This assignment may make $h_j$ full ($a_j = c_j$) or oversubscribed ($a_j > c_j$). In both cases, all \emph{strict} successors of $h_j$'s $c_j^{th}$ assignee are then removed from its preference list and it from theirs. This deletion may make some resident $r_k$ previously assigned to $h_j$ to be free again (if this happens $a_j$ is decremented). That resident can then apply to the next hospital on his/her preference list. The process will continue while some resident is free and has a hospital on his/her preference list. 

Other steps were taken to improve the optimisation performance of the models. These include placing a lower bound on the objective function and providing an initial solution to the CPLEX solver. Both can be obtained by executing any of the approximation algorithms \cite{IM09} on the HRT instance (the $3/2$-approximation algorithm for HRT with ties on one side only due to Kir\'{a}ly \cite{Kir08} was chosen). 

\subsection{Testing the implementation}
Although the theoretical model has been proven to be correct, it is still important to verify the correctness of the implementation. The system was tested to ensure a high degree of confidence in the results obtained. The correctness of the pre-processing steps and the IP solution were evaluated by generating multiple instances (100,000) of various sizes (with up to 400 residents) and testing the stability and size of the resulting matching against both the original and the trimmed problem instance. The matchings generated can also be tested to ensure that the hospitals do not exceed their capacities, residents are assigned to at most one hospital, and assignments are consistent (meaning that if $r_i$ is assigned to $h_j$ then $h_j$ is also assigned to $r_i$). For all the instances tested, the solver produced stable matchings. 

Stability is just one property of the resulting matching that needs to be tested. Another important property is optimality. One method of testing the optimality of the stable matchings produced would be to compare the sizes of optimal solutions generated by various models on the same problem instance. Although we have presented only one IP model,  HRT can be modeled as integer programs in various ways. Implementing these different models and comparing their outputs gives some indication of the correctness of the implementation. This test is not fool-proof as errors in the implementations may be duplicated in the models. We would have to test for optimality using a more exhaustive approach.

To verify that the stable matchings generated by the IP model are optimal, another method is to generate all stable matchings in the HRT model. The largest stable matching found would then be compared to the optimal solution obtained by the solver. If the sizes of both matchings are the same, we confirm that the IP solver produced an optimal solution to that instance. A difference would indicate a fault in one or both of the methods. The larger the number of instances tested and the bigger the size of instances, the more confident we will be of the IP implementation. The generation of all stable matchings was done by finding all matchings in the underlying bipartite graph, testing for stability of each matching and keeping a record of the largest found. Due to the brute-force nature of this technique we could only generate and test 100 instances each of size 8, 9, 10, 11 and 12 to optimality. For all the instances tested, the solver produced optimal stable matchings.

\section{Empirical evaluation}
\label{eval}
An empirical evaluation of the IP model was carried out. Large numbers of random instances of HRT were generated by varying certain parameters relating to the construction of the instance and passed on to the CPLEX IP solver. Data from past SFAS matching runs were also modelled and solved. This section discusses the methodology used and some of the results obtained. Experiments were carried out on a Linux machine with 8 Intel(R) Xeon(R) CPUs at 2.5GHz and 32GB RAM.

\begin{table}[!h]
\centering
\begin{tabular}{|c|c|c|c|}
\hline
$t_d$ & $n_1 = 200$ & $n_1 = 250$ & $n_1 = 300$ \\ \hline \hline
75$\%$ & 100.00$\%$ &100.00$\%$&99.85$\%$ \\  \hline
80$\%$ & 99.98$\%$ &99.88$\%$&99.39$\%$ \\ \hline
85$\%$ & 99.90$\%$ &99.29$\%$&97.76$\%$ \\ \hline
90$\%$ & 99.70$\%$ &99.28$\%$&98.60$\%$ \\ \hline
95$\%$ & 99.99$\%$ &100.00$\%$&100.00$\%$ \\ \hline
\end{tabular}
\caption{Percentage solvable instances ($100\%$ for omitted $t_d$ values)}
\label{prec}
\end{table}

\subsection{Using random instances}
Random HRT problem instances were generated. The instances consist of $n_1$ residents, $n_2$ hospitals and $C$ posts where $n_1$, $n_2$ and $C$ can be varied. The hospital posts were randomly distributed amongst the hospitals.  Other properties of the generated instance that can be varied include the lengths of residents' preference lists as well as a measure of the density $t_d$ of ties present in the preference lists. The tie density $t_d$ $(0 \leq t_d \leq 1)$ of the preference lists is the probability  that some agent is tied to the agent next to it in a given preference list. At $t_d = 1$ each preference lists would be contained a single tie while at $t_d = 0$ no tie would exist in the preference lists of all agents thus reducing the problem to an HR instance. We define the size of the instance as the number of residents 
$n_1$ present.  

\subsubsection{Varying tie density}
Since ties cause the size of stable matchings to vary, an obvious question to investigate is how the variation in tie density affects the runtime of the IP model and the size of the maximum stable matchings found. 
These values were measured for multiple instances of MAX HRT while varying the tie density $t_d$ of hospitals' preference lists. There were no ties in the residents' preference lists. This was done for increasing sizes ($n_1=200, 250, 300$) of the problem instance with the residents' preference lists  being kept strictly ordered at 5 hospitals each. A total of 10000 instances were randomly generated for each tie density value (starting at $t_d=0\%$ to $t_d=100\%$ with an interval of $5\%$) and instance size. For each instance $C=n_1$ and $n_2 = \lfloor0.07 \times nR\rfloor$.

To avoid extreme outliers skewing the mean measures, we define what we regard as a reasonable solution time (300 seconds) and abandon search if the solver exceeds this cut-off time. For most tie densities this cut-off was not exceeded for the values of $n_1$ and $t_d$ considered (omitted $t_d$ values were $100\%$ solvable). Table \ref{prec} shows the percentage of instances that were solved before the cut-off was exceeded. 

\begin{figure}[h]
\centering
\begin{minipage}[b]{0.45\linewidth}
\centering
\setlength\fboxsep{0pt}
\setlength\fboxrule{0pt}
\fbox{\includegraphics[width=200pt]{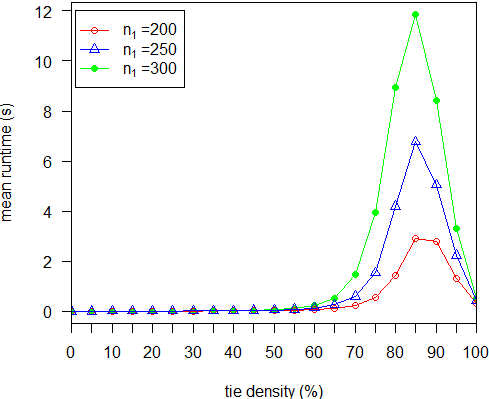}}
\caption{Mean runtime vs $t_d$}
\label{vary_tie_mean}
\end{minipage}
\hspace{0.5cm}
\begin{minipage}[b]{0.45\linewidth}
\centering
\setlength\fboxsep{0pt}
\setlength\fboxrule{0pt}
\fbox{\includegraphics[width=200pt]{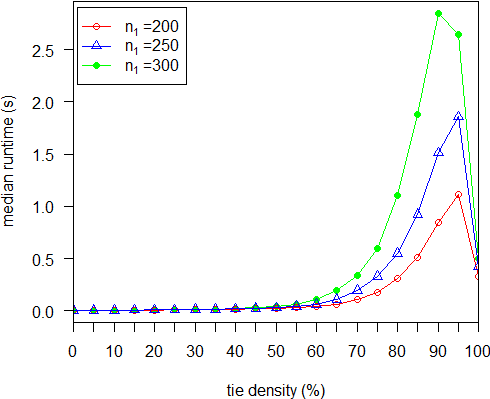}}
\caption{Median runtime vs $t_d$}
\label{vary_tie_median}
\end{minipage}
\end{figure}

From Figures \ref{vary_tie_mean} and \ref{vary_tie_median} we see that the mean and median runtime remain significantly low for instances with $t_d < 60\%$ but then gradually increase until they reach their peaks (in the region of $80\%-90\%$) before falling as the tie density approaches $100\%$. From a theoretical perspective, it is known that the problem is polynomially solvable when the tie density is at both $0\%$ and $100\%$ and it is easy to see how the IP solver will find these cases trivial. As the tie density increases the number of stable matchings that the instance is likely to admit also increases, explaining the observed increase in the runtime. The \emph{hospitals-offer} and \emph{residents-apply} algorithms used to trim the instance also play their part in this trend with limited trimming done for higher tie densities.

\begin{figure}[h]
\centering
\setlength\fboxsep{0pt}
\setlength\fboxrule{0pt}
\fbox{\includegraphics[width=200pt]{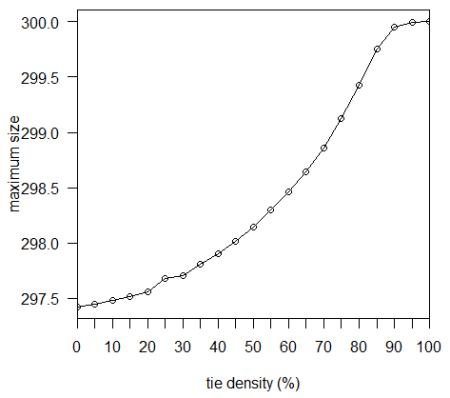}}
\caption{$|M|$  vs $t_d$ for $n_1=300$}
\label{td_size}
\end{figure}

Figure \ref{td_size} shows the variation in optimal values with tie density for $n_1=300$. We observe an increase in the average size of maximum stable matchings as the tie density increases. This is in line with the idea that the stability requirement  restricts the size of stable matchings and increasing tie density can be viewed as relaxing stability requirements.  

\subsubsection{Increasing instance size}
The execution time for solving multiple MAX HRT IP model instances of increasing sizes was evaluated. The tie density $t_d$ and preference list lengths were kept constant. This provided an estimate of problem sizes for which the IP model can be of practical value. The tie densities of the hospitals' preference lists were set to 0.85 on all instances. On the bases of the results from the previous section, we estimate that instances with tie density in this region (0.7--0.95) will take longer to be solved than others. There were no ties on the residents' preference lists. The instance size $n_1$ was increased  by 50 starting at $n_1=100$. A total of 100 instances for each instance size was generated. The number of hospitals $n_2$ in each instance was set to $\lfloor 0.07 \times n_1 \rfloor$. Each resident has a preference list of 5 hospitals with each hospitals' preference list length being determined by the frequency of its occurrence within the residents' lists. 

\begin{figure}[h]
\centering
\begin{minipage}[b]{0.45\linewidth}
\centering
\setlength\fboxsep{0pt}
\setlength\fboxrule{0pt}
\fbox{\includegraphics[width=200pt]{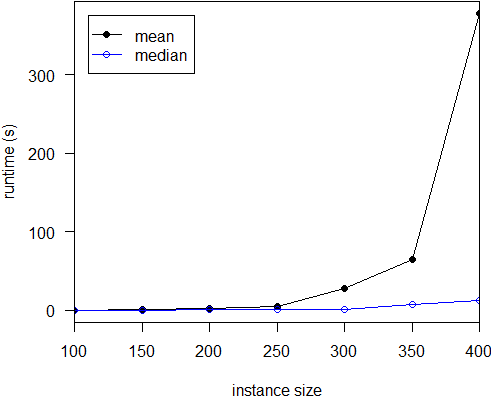}}
\caption{Mean  and median runtime vs instance size}
\label{inc_size_time}
\end{minipage}
\hspace{0.5cm}
\begin{minipage}[b]{0.45\linewidth}
\centering
\setlength\fboxsep{0pt}
\setlength\fboxrule{0pt}
\fbox{\includegraphics[width=200pt]{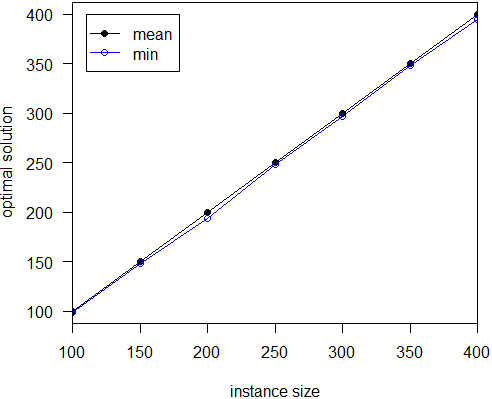}}
\caption{Optimal solution size vs instance size}
\label{inc_size_size}
\end{minipage}
\end{figure}

Table \ref{inc_size_time} shows how the mean and median runtime rise as $n_1$ increases. We assume the sharp difference between the mean and median is due to the presence of outliers corresponding to exceptionally difficult instances.  As the instance size grows the existence of challenging outliers becomes more probable, leading to a more rapid increase in the mean runtime compared to the median runtime. Table \ref{inc_size_size} shows the difference between the mean and min optimal solutions found. It is interesting to observe that optimal solutions tend to match almost every resident given the instance generation parameters used. 


\subsection{Using real instances}
Another question worth asking is whether the IP model can handle instance sizes found in real-world applications. In \cite{IM09}, various approximation algorithms and heuristics were implemented and tested on real datasets from the SFAS matching scheme for 2006, 2007 and 2008 where the residents' preferences are strictly ordered with ties existing in the hospitals' preference lists. With the IP model, it is now possible to trim the instances using the techniques mentioned in Section 3, generate an optimal solution and compare the results obtained with those reported in \cite{IM09}. Results from these tests showed that, while some algorithms did marginally better than others, all the algorithms developed generated relatively large stable matchings with respect to the optimal values. Table \ref{sfas} shows this comparison. Let $M'$ denote the largest stable matching found over all the algorithms tested in \cite{IM09}.

\begin{table}[h]
\centering
\begin{tabular}{|c|c|c|c|c|c|c|}
\hline
year & $n_1$ & $n_2$ & $t_d$ & time (sec) & $|M|$ &$|M'|$ from \cite{IM09} \\ \hline \hline
2006 & 759 &53&92$\%$&92.96&758&754 \\  \hline
2007 & 781 &53& 76$\%$&21.78&746&744 \\ \hline
2008 & 748 &52& 81$\%$&75.50&709&705\\ \hline
\end{tabular}
\caption{SFAS IP Results}
\label{sfas}
\end{table}

\section{Future work}
\label{future}
It still remains to be tested whether various node, variable and value ordering heuristics would help improve the performance of the IP model. Another interesting idea is to investigate the possibility of adopting a column generation technique for the IP model in order to improve scalability so that larger instance sizes can be solved.

\section*{Acknowledgments}
We would like to thank Iain McBride for useful discussion concerning the IP model presented in Section 3.
\bibliography{../../matching_db}
\end{document}